# Astro2020 Science White Paper

**The Importance of 3D General Circulation Models for Characterizing the Climate and Habitability of Terrestrial Extrasolar Planets.**

**Thematic Areas:** Planetary Systems

**Principal Author:**
Name:          Eric. T. Wolf
Institution:  University of Colorado, Boulder,
Email:          eric.wolf@colorado.edu
Phone:        240 - 461 - 8336

**Co-authors:** Ravi Kopparapu (NASA GSFC), Vladimir Airapetian (NASA GSFC & American University), Thomas Fauchez (NASA GSFC & USRA), Scott D. Guzewich (NASA GSFC), Stephen R. Kane (UC Riverside), Daria Pidhorodetska (NASA GSFC), Michael J. Way (NASA GISS)

**Co-signers:** Dorian S. Abbot (University of Chicago), Jade H. Checlair (University of Chicago), Christopher E. Davis (University of Washington), Anthony Del Genio (NASA GISS), Chuanfei Dong (Princeton University), Siegfried Eggl (University of Washington), David P. Fleming (University of Washington), Yuka Fujii (ELSI), Nader Haghighipour (IfA, Hawaii), Nicholas Heavens (Hampton University), Wade G. Henning (University of Maryland, NASA GSFC), Nancy Y. Kiang (NASA GISS), Mercedes López-Morales (Harvard-Smithsonian Center for Astrophysics), Jacob Lustig-Yaeger (University of Washington), Vikki Meadows (University of Washington), Christopher T. Reinhard (Georgia Institute of Technology), Sarah Rugheimer (Oxford), Edward W. Schwieterman (University of California, Riverside), Aomawa L. Shields (University of California, Irvine), Linda Sohl (Columbia University), Martin Turbet (Laboratoire de Meteorologie Dynamique/IPSL, CNRS), Robin. D. Wordsworth (Harvard University)

**Summary**

While recently discovered exotic new planet-types have both challenged our imaginations and broadened our knowledge of planetary system workings, perhaps the most compelling objective of exoplanet science is to detect and characterize habitable and possibly inhabited worlds orbiting in other star systems.  Our technological capabilities currently preclude humans from visiting or sending probes to extrasolar planets for in situ explorations.  Thus, for the foreseeable future, characterizations of extrasolar planets will be made via remote sensing of planetary spectroscopic and temporal signals, along with careful fitting of this data to advanced models of planets and their atmospheres.  Terrestrial planets are small and significantly more challenging to observe compared to their larger gaseous brethren; however observatories coming on-line in the coming decade will begin to allow their characterization.  Still, it is not enough to invest only in observational endeavors.  Comprehensive modeling of planetary atmospheres is required in order to fully understand what it is that our grand telescopes see in the night-sky.  In our quest to characterize habitable, and possibly inhabited worlds, 3D general circulation models (GCMs) should be used to evaluate potential climate states and their associated temporal and spatial dependent observable signals.  3D models allow for coupled, self-consistent, multi-dimensional simulations, which can realistically simulate the climates of terrestrial extrasolar planets.  A complete theoretical understanding of terrestrial exoplanetary atmospheres, gained through comprehensive 3D modeling, is critical for interpreting spectra of exoplanets taken from current and planned instruments, and is critical for designing future missions that aim to measure spectra of potentially habitable exoplanets as one of their key science goals.  We recommend continued institutional support for 3D GCM modeling teams that focus on planetary and exoplanetary applications.

**Introduction**

Recently, terrestrial sized planets have been detected in the habitable zones of a variety of host stars, such as Proxima Centauri b (Anglada-Escudé et al. 2016), LHS 1140b (Dittman et al. 2017), Kepler 62f (Borucki et al. 2013), GJ 1132b (Berta-Thompson et al. 2015), Kepler 442b (Torres et al. 2015), Ross 128b (Bonfils et al. 2018), and the TRAPPIST-1 system (Gillon et al. 2017).  Transmission spectroscopy of the TRAPPIST-1 planets between 1.1 and 1.7 µm has already been attempted with the *Hubble Space Telescope* (de Wit et al. 2018).  Detailed transmission spectroscopy across the near-infrared is within reach of the *James Webb Space Telescope* for numerous known terrestrial sized exoplanets (Morley et al. 2017).  Furthermore, the *Transiting Exoplanet Survey Satellite* is expected to identify many new terrestrial planets in the habitable zones of nearby stars which will be excellent candidates for follow up observations (Ricker et al. 2015).  Upcoming ground-based telescopes such as the *European Extremely Large Telescope* and the *Thirty Meter Telescope* should be able to directly image Proxima b (Turbet et al. 2016; Lovis et al. 2017).  Future space telescopes that are currently being studied, such as *LUVOIR*, *HabEX*, *and OST*, are being specifically designed for characterizing the atmospheres of terrestrial extrasolar planets in the habitable zone.  In the next decade retrieving spectra from terrestrial planets in the habitable zones of low-mass star may be within reach.  Within the next two decades, we will likely be able to retrieve spectra from terrestrial planets in the habitable zones of Sun-like stars.  The day is close at hand where we will retrieve high quality observations of habitable extrasolar worlds around G, K, and M stellar spectral types.  However, our observations will only be as good as the models used to interpret them.

Extensive modeling of extrasolar planetary climate and habitability has been conducted using energy balance (e.g. Budyko 1969; Hart 1979) and 1D radiative-convective models (e.g. Kasting et al. 1993; Selsis et al. 2007; Kopparapu et al. 2013; Kopparapu et al. 2014; Ramirez & Kaltenegger 2017; 2018) to define the boundaries to the so-called habitable zone, i.e. the region in space where, given an appropriate atmospheric composition, a terrestrial planet can feasibly maintain abundant surface liquid water.  These works have shaped our thinking with regard to the limits of planetary habitability; however these lower dimensional models miss important interconnections within planetary climate systems, which are important for determining the habitability and detectability of any specific target.  Atmospheric processes are inherently three-dimensional.  Atmospheres are spherical shells of fluid that envelop a planet, with motions driven by radiation from the host star and the bulk rotation of the planet.  The climate of planets depends on many things, including the total stellar irradiation, the atmospheric composition, the volatile inventory, the planetary rotation rate, the surface properties, and the presence of clouds or hazes (Forget & Leconte, 2015).  In the past decade, 3D general circulation models (GCMs) have become commonly used to study terrestrial planets in or near the habitable zone (e.g. Wordsworth et al. 2011; Abe et al. 2011; Pierrehumbert 2011; Selsis et al. 2011; Leconte et al. 2013; Wang et al. 2014; Yang et al. 2013, 2014; Shields et al. 2013, 2016; Koll & Abbot 2015; Wordsworth et al. 2015; Pierrehumbert & Ding, 2016; Bolmont et al. 2016; Way et al. 2017, 2018; Popp et al. 2016; Wolf et al. 2017; Kopparapu et al. 2016; 2017; Turbet et al. 2016; 2017; Boutle et al. 2017; Fujii et al. 2017; Haqq-Misra et al. 2018; Kodama et al. 2018; Bin et al. 2018; Del Genio et al. 2019; Kane et al. 2019).  GCMs are able to couple many processes together in a three-dimensional time-marching calculation.

**3D Models of Climate**

A 3D climate model consists of a series of columns of global coverage, interconnected by a dynamical core.  Within each column, physics along the vertical dimension are prognostically calculated, including stellar and thermal radiative transfer, convection, turbulence, cloud condensation, precipitation, evaporation, chemistry, aerosols, and surface processes.  The physical parameterizations used can significantly vary in complexity between models.  The dynamical core solves the Navier-Stokes equations for fluid dynamics on a rotating sphere, and thus predicts the general circulation of the atmosphere.  The dynamical core links the model columns together through horizontal transports of energy and atmospheric constituents by three-dimensional wind fields.   The dynamical core lies at the heart of any GCM.  When a GCM employs detailed physical processes, such as ocean, ice-sheet, and land surface models, for specific representation of Earth-like planets, they are often also called climate system models or Earth system models.

The inherent strength of using 3D GCMs is that they allow for a self-consistent and coupled treatment of all physical process occurring in a planetary atmosphere.  Of particular importance for habitable worlds, is the treatment of water in its various thermodynamic phases throughout the climate system. By definition habitable worlds must have abundant surface liquid water. However, water is responsible for both the sea-ice albedo and water-vapor greenhouse feedbacks which are both strong and positive, meaning that they amplify temperature perturbations to the climate system in either direction. Both sea-ice and water vapor are temporally and spatially variable components to the climate system. Sea-ice and snow deposits are governed by horizontal gradients in temperature and meteorology respectively, both strongly influencing the albedo. The interaction between water vapor and atmospheric dynamics controls the relative humidity of the atmosphere, and ultimately the strength of a planet's water vapor greenhouse effect (Pierrehumbert 1994). While cloud modeling remains notoriously challenging due to the complexity of sub-grid scale and microphysical process, 3D models allow for the spatial resolution of clouds based on fundamental properties of moisture and temperature. Atmospheric circulations play a significant role in determining where cloud condensation will take place, on a three-dimensional world. Interactions between water, ice, clouds, atmospheric circulations, and 3D temperature distributions cannot be prognostically captured by lower dimensional models. Earth-centric tunings of relative humidity and surface albedo can be applied to lower dimensional models, but they are unlikely to remain valid as planet properties differ for extrasolar planets. Figure 1 shows low, middle, and high cloud coverages predicted from 3D coupled atmosphere-ocean simulations of Proxima b, assuming a completely ocean covered planet (Del Genio et al. 2019).

     Of particular concern, 1D models struggle to simulate planets within the habitable zones of low-mass stars (late-K and all M-dwarfs), and it is precisely these planets which will be amenable for characterization due to their favorable system configurations (*i.e.* these host stars are small and have low luminosities, and habitable zone planets have relatively short orbital periods). Terrestrial planets in the habitable zones of low-mass stars are expected to be tidally locked into synchronous rotation (Leconte et al. 2015; Barnes 2017), where one side of the planet is permanently bathed in sunlight, and the other side lies in permanent darkness. Furthermore, atmospheric circulation patterns are strongly controlled by the Coriolis force, which itself is determined by the planetary rotation rate. For synchronous rotating planets, the planet's rotational period equals the orbital period, thus providing a significant constraint for use in our planetary models. Synchronously rotating planets in the habitable zone generally have rotation periods that are much longer than the Earth, meaning that the Coriolis force is much weaker. This has a profound effect on the atmospheric circulation (Joshi 2003; Merlis & Schneider 2011), changing the circulation state from primarily zonal flow with mid-latitude jets, to sub-stellar to

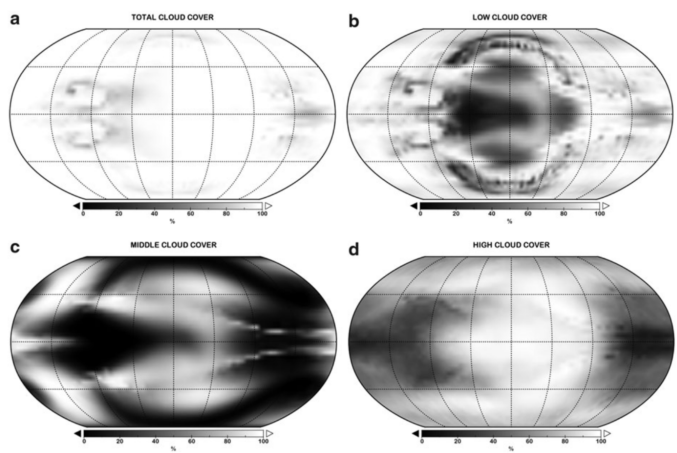

Figure 1: Total (a), low (p > 680 mb) (b), middle (440 < p < 680 mb) (c), and high (p < 440 hPa) (d) cloud cover, from 3D simulations of Proxima b, from Del Genio et al. (2019).

anti-stellar radial flow aloft with strong rising motions on the permanent day-side, and subsiding motions on the permanent night-side.

These changes to the circulation state, and pattern of stellar instellation (incident flux) with respect to the planet surface, have profound effects on the climate system, which can only be captured by 3D climate system models. Yang et al. (2013) showed that slowly and synchronously rotating planets have thick clouds that encompass the substellar hemisphere, significantly increasing the planet's albedo, and allowing the planets to avoid entering a runaway greenhouse at significantly higher incident stellar fluxes compared to an Earth-twin. Kopparapu et al. (2016) showed the ubiquitous substellar cloud deck for synchronous rotators is sensitive to the specific rotation period (equal to the orbital period) of the planet. For faster rotating planets, like TRAPPIST-1e with only a 6.1 day rotation period, the Coriolis force grows strong enough to advect clouds eastward off of the substellar point, reducing their efficacy at increasing the planet's albedo. Other researchers have highlighted the importance of correctly treating dynamic ocean heat transport for more accurately simulating the climates of slowly rotating habitable planets (Hu & Yang, 2014; Del Genio et al. 2019; Way et al. 2018; Yang et al. 2019). Recently, 3D models have been used to study the coupled climate-chemistry of Earth-like extrasolar planets, and have found day-night chemical gradients can exist on slow rotating planets (Chen et al. 2018). Finally, still others have found that synchronously rotating planets may be much more prone to cold trapping of volatile species on their cold permanent night-sides (Leconte et al. 2013; Turbet et al. 2017). For simulating synchronously rotating planets around low-mass stars, there are numerous physical processes which require three-dimensions in order to adequately assess them.

**Relevance to Observations**

The most simplistic characterizations of exoplanet habitability can be made by comparing the total stellar instellation received by an observed exoplanet to theoretically determined habitable zone boundaries. 3D models have already proven valuable in refining habitable zone boundaries (e.g. Leconte et al. 2013; Yang et al. 2014; Wolf et al. 2017; Kopparapu et al. 2017; Turbet et al. 2017b; Paradise & Menou 2017). However, future observatories will do much more than just determine an exoplanet's radius and semi-major axis. Future observatories have the capability to deduce transmission spectra, thermal emission spectra, reflectance spectra, and also their respective variability as a function of orbital phase. The observations needed for true exoplanet characterizations of climate and habitability are inherently multi-dimensional, relying on specific geometries of the star-planet-observer grouping. For instance, transit spectroscopy probes planetary atmospheres only along the terminator region (*i.e.* the division between night and day). The use of global mean profiles from 1D models will not accurately reflect conditions found at the terminator regions, particularly for slowly rotating planets around low-mass stars. Thermal emission spectra are dependent on the radiating level of the atmosphere, and are strongly modulated by the spatial distribution of upper level clouds (Yang et al. 2014; Haqq-Misra et al. 2018). Reflectance spectra are strongly dependent on the distributions of land vs. ocean, sea ice, snow, and cloud cover. Phase and time-dependent observations rely on longitudinal differences in the planet emission and reflection (Cowan et al., 2009; Lustig-Yaeger et al., 2018), which are not possible to capture adequately with energy balance and 1D radiative-convective models.

3D GCMs of hot Jupiters have already proven critical for explaining observations. Notably, strong equatorial superrotation of hot Jovian atmospheres is responsible for the eastward shift from the substellar point of the observed thermal emission maximum (e.g. Showman & Guillot, 2002; Knutson et al. 2007). Such multi-dimensional structure will also prove critical for explaining future observations of habitable terrestrial planets, and initial work supports this inference. For instance, Yang et al. (2013) using a 3D model determined that the thick substellar cloud decks found for slowly rotating planets will lead one to observe a thermal emission minimum at secondary eclipse, due to the strong greenhouse effect of high ice clouds. Conversely, when viewed at transit, such worlds will exhibit a thermal emission maximum due to clear-sky conditions on the night-side, which allows thermal radiation from the lower atmosphere to escape directly to space. Others, have found that the specific rotation period of synchronously rotating habitable planets can cause phase offsets in the thermal emission minimum and maximum, due to the specific locations of clouds in the atmosphere. Figure 2 illustrates the differences in broadband thermal emission phase curves for synchronously rotating temperate ocean-covered planets at various orbital periods (Haqq-Misra et al. 2018). Kopparapu et al. (2017) and Fujii et al. (2017) both showed that the strong upwelling motions at the substellar point cause habitable planets to have significantly more water vapor in their stratospheres, which deepens water vapor signals in transit spectra, compared to an Earth-twin. All of these interesting and observationally relevant features of habitable worlds were determined through use of advanced 3D general circulation models.

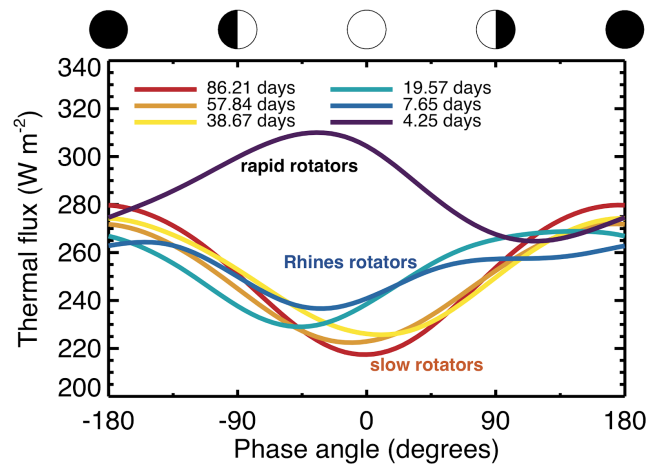

Figure 2: Thermal emission phase curves from 3D simulations of habitable planets orbiting low mass stars, from Haqq-Misra et al. (2018).

**Summary and Recommendation**

- A comprehensive understanding of planetary atmospheres is critical, if scientists are to feel confident in interpreting spectra impartially for convincing indicators of habitability and inhabitance.
- Three-dimensional general circulation models provide a powerful tool in our efforts to understand, identify, and characterize habitable extrasolar planets and their potential observable signals that are otherwise not predicted by lower dimensional models.
- General circulation models are complex codes, that require diverse expertise, teamwork, and community support. *We recommend continued institutional support for 3D GCM modeling teams that focus on planetary and exoplanetary applications.*